\documentstyle[epsf,twocolumn,prl,aps]{revtex}
\begin{document}                                                
\draft
\title{Critical Line for $H$-Superfluidity in Strange Quark Matter?}
\author{Robert D. Pisarski}
\bigskip
\address{
Department of Physics, Brookhaven National Laboratory,
Upton, New York 11973-5000, USA\\
email: pisarski@bnl.gov\\
}
\date{\today}
\maketitle
\begin{abstract} 
Based upon the analogy to the electroweak phase diagram,
I propose that in QCD
there might be a critical line for a superfluid transition,
in the plane of chemical potential and temperature.
The order parameter has the quantum numbers of the $H$-dibaryon,
but the transition is driven by color superconductivity in
strange quark matter.
\end{abstract}

In QCD, there is a phase transition to a color superconducting
phase at high quark density and low temperature 
\cite{bl,gen,rg,cfl,con,prl,prs,son,sch,gap,rock,qcd,cdw,q3a,q3b,twoa,twob,abr,rssv,proc,cas,sonst,ehhs,dirk}.  
At densities
of interest for the collisions of heavy ions or quark stars, 
``$2+1$'' flavors of quarks --- up, down, and strange --- enter.

The order of the phase transition to a color superconducting phase
at zero temperature, as function of the quark chemical potential,
was analyzed previously \cite{prl}.  
The zero temperature transition is simple because 
the effective theory is four dimensional over large distances \cite{hsu4}.
For a second order transition, couplings can only flow into the
origin, with mean field behavior corrected by logarithms.  
Most importantly, quark loops screen gluons,
so that gluons do not contribute over long distances.  
For $2+1$ flavors, this analysis predicts
a first order transition \cite{prs}.

The transition at nonzero temperature is much more complicated.
Over large distances, the effective theory is three dimensional;
a second order transition typically flows toward a fixed point 
which lies in a regime of strong coupling.  Also, while 
static electric fields are screened by quark loops, static magnetic fields
are not.  Thus the phase transition involves scalar fields coupled
to gauge fields in three dimensions.

In this paper I consider the effective 
theories which are of relevance for the phase transition to
color superconductivity for $2+1$ flavors of quarks \cite{prl}.
This enables me to unify a large number of model dependent
results in a simple manner.  Because of an instanton induced
term \cite{cfl,q3a}, I find one surprise.  As suggested previously
\cite{bl,rg,cfl,prl,rssv,ehhs}, in a chirally 
symmetric, color superconducting phase, 
an (approximate) spontaneous violation
of parity can be large.  The pattern, however, is
unexpected: {\it if} instantons are important, then
(approximate) parity violation is greater for the 
up-strange and the down-strange superconducting condensates
than it is for the up-down condensate.

While the phase transitions of scalars coupled to gauge fields in
three dimensions is a complicated problem, 
because of the possibility of generating a cosmological baryon
asymmetry at the electroweak scale, 
much is known about such phase diagrams 
from numerical simulations on the lattice \cite{ew,adj}.
Using this information, 
I conjecture what the phase diagrams for the effective
three dimensional theories for color superconductivity might look like.
Following especially the phase diagram for adjoint scalars coupled
to a $SU(3)$ gauge field \cite{adj},
I suggest that for $2+1$ flavors, 
there might be a {\it line} of second 
order phase transitions, in the plane of chemical
potential and temperature.  
The transition is induced by color superconductivity for $2+1$ flavors, 
assuming color-flavor locking \cite{cfl}.  Even so, it is 
properly a superfluid transition,
where the order parameter is an operator
for the $H$-dibaryon \cite{con,prl}.
Like ordinary superfluidity, ``$H$-superfluidity''
lies in the universality class of an $O(2)$ vector.

All of my arguments are qualitative and, on occassion, speculative.
However, the phase diagram
for the effective three dimensional theory is directly testable
by lattice simulations involving only scalars and gauge fields.
Where the critical line for $H$-superfluidity begins (if at all)
can be then estimated by using perturbation theory in QCD.
If a critical line does occur, however, it is manifestly of
experimental interest, as is 
a critical end point for the chiral phase transition \cite{rss}.

\section{Effective theories}

In this section 
I first review the order parameters for color superconductivity
with massless quarks
\cite{bl,prl,prs,gap}, and then use them to construct
effective lagrangians in a standard fashion.
I assume that if a condensate 
with (total) spin zero can form --- as is true for two and three flavors ---
that it does, and dominates over condensates with
higher spin.  

Massless quarks naturally decompose into eigenstates of chirality.
In a Fermi sea, particles have zero energy
near the Fermi surface, and dominate over anti-particles, which always
have nonzero energy.  This it is natural to introduce 
projectors for chirality and energy,
\begin{equation}
{\cal P}_{R,L} \, = \,
\frac{1}{2} \left( 1 \pm \gamma_5\right) 
\;\;\; , \;\;\;
{\cal P}^{\pm} \, = \, \frac{1}{2} \left( 1 \pm \gamma_0
{\bf \gamma} \cdot {\hat k} \right)  \; ,
\label{e1}
\end{equation}
where $\vec{k}$ is the momentum of the quark, and $\vec{k} = k \hat{k}$,
$\hat{k}^2 = 1$.  There are then four types of quark fields, right
and left handed, and particle and anti-particle.

Quarks transform under a local gauge group of $SU(3)_c$ color;
the color indices of 
the fundamental representation are denoted by $i,j=1,2,3$.
For $N_f$ flavors of massless quarks, 
with flavor indices $a,b=1...N_f$, classically there is also a 
global flavor symmetry of
$SU(N_f)_R \times SU(N_f)_L \times U_A(1) \times U(1)$.
A right handed particle is given by
\begin{equation}
q^+_{i,a_R} \; = \; {\cal P}^+ \, {\cal P}_{R} \, q_{i,a} \; ,
\label{e2}
\end{equation}
where $q_{i,a}$ is a quark field with color $i$, flavor $a$,
and momentum $\vec{k}$.

There are two right handed superconducting condensates with total
spin zero: between two right handed particles, or two 
right handed anti-particles,
\begin{equation}
\Phi^{\pm}_{i,j;a_R, b_R} \; = \;
(q^{\pm}_{i,a_R})^T \; C \; q^{\pm}_{j,b_R} \; ,
\label{e3}
\end{equation}
and similarly for the left handed condensates.  
$q$ has momentum $\vec{k}$,
$q^T$ is the Dirac transpose of a quark with momentum $- \vec{k}$,
and $C$ is the charge conjugation matrix.  Gaps for (total)
spin one are constructed similarly \cite{gap}.

Superconductivity is due to pairing of particles near the Fermi
surface, so it is natural to expect that only the 
particle condensates, $\Phi^+$, matter, and
that the anti-particle condensates, $\Phi^-$, can be
neglected.  In an effective lagrangian approach, this happens as follows.  
As is evident from (\ref{e3}), for every particle
condensate there is a corresponding anti-particle condensate.
Thus in an effective lagrangian the two fields mix,
\begin{equation}
g^2 \; {\rm tr}\left(
(\Phi^-)^\dagger \Phi^+ \; + \; {\rm c.c.} \right)
\; + \; m_-^2 \; {\rm tr} \left( | \Phi^-|^2 \right) \; .
\label{e3a}
\end{equation}
I assume that $\Phi^-$ does not condense on its own, so that
it has a positive mass squared, $m_-^2 > 0$.  For free fields,
$\Phi^+$ and $\Phi^-$ do not mix, but they do at $\sim g^2$,
since interactions invariably mix particles and anti-particles \cite{dirk}.
Here $g$ is the QCD coupling constant, although perhaps
the $g^2$ is only a $g$, due to a logarithmic enhancement
from forward scattering \cite{son,sch,gap,rock,qcd,cdw,q3a,q3b,dirk}.
Whatever the value of the mixing term, though, excluding isolated
points in the phase diagram, there is no generic reason why
it should vanish.  With (\ref{e3a}), when $\Phi^+$ condenses,
it becomes a term linear in $\Phi^-$, so it also condenses,
$\langle \Phi^- \rangle \sim g^2 \langle \Phi^+ \rangle$.  
But the critical behavior, where $\langle \Phi^+ \rangle \rightarrow 0$,
is dominated by $\Phi^+$ alone.  
Thus I consider only the particle condensates and drop the
``$+$'' superscript, $\Phi = \Phi^+$.

Besides those for color superconductivity,
I also require the order parameters for chiral symmetry breaking.
Chiral symmetry is broken by a condensate between an anti-quark and
a quark.  From group theory, the product of a color anti-triplet
and a triplet is a singlet plus an octet,
$\overline{{\bf 3}} \times {\bf 3} = {\bf 1} + {\bf 8}$.
There are then two chiral order parameters: a color singlet,
\begin{equation}
\psi_{a_L,b_R} \; = \;
\overline{q}_{i,a_L} \, q^{}_{i,b_R} \; ,
\label{e4}
\end{equation}
and a color adjoint field, 
\begin{equation}
\widetilde{\psi}^C_{a_L,b_R} \; = \;
\overline{q}_{i,a_L} \; t^C_{ij} \; q^{}_{j,b_R} \; ;
\label{e4a}
\end{equation}
$t^C_{ij}$ is the generator for $SU(3)_c$, with
the adjoint index $C = 1\ldots 8$.
In the vacuum, 
\begin{equation}
\langle \overline{q}_{i,a_L} \, q_{j,b_R} \rangle
\; = \; \psi_0 \;  \frac{\delta_{i j}}{3} \; \delta_{a_L b_R} \; ;
\label{e4b}
\end{equation}
this leaves $SU(3)_c$ unbroken, and breaks
the flavor $SU(N_f)_R \times SU(N_f)_L \rightarrow SU(N_f)$.
The color singlet chiral field develops an expectation
value, $\langle \psi \rangle \sim \psi_0$, and the
color adjoint chiral field does not, 
$\langle \widetilde{\psi} \rangle = 0$.  

Using this inelegant notation, one can write 
down how the fields 
transform under the nonabelian symmetries.
What is simpler and more useful is how they transform under
the abelian flavor symmetries of $U_A(1) \times U(1)$.  
Suppressing the color and flavor indices, the quark
fields transform as 
\begin{equation}
q_R \, \rightarrow \,
{\rm e}^{i (\theta + \theta_A)} \, q_R \;\;\; , \;\;\;
q_L \, \rightarrow \,
{\rm e}^{i (\theta - \theta_A)} \, q_L \; ,
\end{equation}
so the condensate fields transform as
$$
\Phi_R \; \rightarrow \; 
{\rm e}^{2 i (\theta + \theta_A)} \; \Phi_R \;\;\; ,\;\;\;
\Phi_L \; \rightarrow \; 
{\rm e}^{2 i (\theta - \theta_A)} \; \Phi_L \; ,
$$
\begin{equation}
\psi \; \rightarrow \;
{\rm e}^{2 i \theta_A} \, \psi \;\;\; , \;\;\;
\widetilde{\psi} \; \rightarrow \;
{\rm e}^{2 i \theta_A} \, \widetilde{\psi} \; .
\label{e5}
\end{equation}
$\theta$ generates rotations for the $U(1)$ symmetry of quark number,
which is an exact symmetry of the lagrangian.
In contrast, $\theta_A$ generates a rotation for the  
$U_A(1)$ symmetry of anomalous quark number; this is
badly broken in the vacuum, but at high density or temperature, is
very nearly a good symmetry of the lagrangian \cite{hooft}.
Note that $\Phi_{R,L}$ transform nontrivially under both
$U_A(1)$ and $U(1)$, while $\psi$ 
and $\widetilde{\psi}$ transform only under
the anomalous $U_A(1)$.

Color superconductivity involves quarks pairing with quarks,
so Fermi statistics implies a nontrivial relation.  
For a spin zero condensate, $\Phi$ must be
symmetric in the simultaneous exchange of color and flavor indices,
\begin{equation}
\Phi_{R,L}^T \; = \; 
+ \, \Phi^{}_{R,L} \; .
\label{e6}
\end{equation}
(Condensates with spin one satisfy a more complicated
relationship, but are essentially antisymmetric \cite{gap}.)
There is no such relationship for chiral symmetry breaking,
which involves the condensation of quarks with anti-quarks.

Group theory tells us that the product of two color triplets is an
anti-triplet plus a sextet,
${\bf 3} \times {\bf 3} = \overline{{\bf 3}}_a + {\bf 6}_s$;
the subscripts denote anti-symmetric and symmetric representations,
respectively.
By (\ref{e6}), the color anti-triplet piece of $\Phi$, which I denote $\phi$,
combines with an anti-symmetric
flavor representation, while the color sextet part, $\chi$,
combines with a symmetric flavor representation.
Under singlet gluon exchange, the anti-triplet channel
is attractive, and the sextet repulsive.

Defering the precise definitions of $\phi$ and $\chi$ for now, 
the lowest order effective lagrangian, including
gauge interactions, but neglecting terms which are nonlinear in the
condensate fields, is
\begin{equation}
{\cal L}^0 \; = 
\; {\cal L}_\psi^0 
\, + \, {\cal L}_{\widetilde{\psi}}^0
\, + \, {\cal L}_\phi^0 
\, + \, {\cal L}_\chi^0 
\, + \, {\cal L}_\psi^m
\, + \, {\cal L}_g \; . 
\label{e7}
\end{equation}
For massless quarks, the effective lagrangian is composed of
four terms: for the color singlet chiral field,
\begin{equation}
{\cal L}_\psi^0 \; = \; {\rm tr}\left(|\partial_\alpha \psi|^2 \right)
\, + \,
m_\psi^2 {\rm tr}\left( \psi^\dagger \psi \right) \; ,
\label{e8}
\end{equation}
the color adjoint chiral field, 
\begin{equation}
{\cal L}_{\widetilde{\psi}}^0
\; = \; {\rm tr}\left(|D_\alpha \widetilde{\psi}|^2
\right) \; + \;
m_{\widetilde{\psi}}^2 \; 
{\rm tr}\left(\widetilde{\psi}^\dagger \widetilde{\psi}\right) \; ,
\label{e8a}
\end{equation}
the color anti-triplet superconducting field,
\begin{equation}
{\cal L}_\phi^0 \; = \; 
{\rm tr}\left(|D_\alpha \phi|^2 \right)
\; + \; m_\phi^2 \; {\rm tr}\left(\phi^\dagger \phi\right) \; ,
\label{e9}
\end{equation}
and the color sextet superconducting field,
\begin{equation}
{\cal L}_\chi^0 \; = \; {\rm tr}
\left( |D_\alpha \chi|^2 \right)\; + \; 
m_\chi^2 \; {\rm tr}\left(\chi^\dagger \chi\right) \; ;
\label{e10}
\end{equation}
$D_\alpha$ is the covariant color derivative in
the appropriate representation.  For massive quarks, one
also needs
\begin{equation}
{\cal L}_\psi^m \; = \;
+ \, {\rm tr}\left( \psi {\cal M} \right) \; .
\label{e10b}
\end{equation}
The diagonal elements of ${\cal M}$ are
proportional to the current quark masses,
${\cal M}_{a_L b_R} \sim m_a \delta_{a_L b_R} $,
with $m_a$ is the current quark mass for flavor $a$.
From current algebra and lattice simulations \cite{mass}, the 
quark masses for up, down, and strange are
$m_u \sim 4 MeV$, $m_d \sim 8 MeV$, and $m_s \sim 100 MeV$, respectively.

The lagrangian for the color gauge field, ${\cal L}_g$, is the
usual action plus a term for hard dense loops \cite{hdl}.
I assume that the Debye mass for hard dense loops is always nonzero.

While these terms are all completely standard, given the multiplicity
of fields, it helps to be explicit.
I assume that the adjoint chiral field and the color sextet
field always represent repulsive channels, with positive mass squared:
\begin{equation}
m_{\widetilde{\psi}}^2 \; > \; 0 \;\;\; , \;\;\; 
m_\chi^2 \; > \; 0 \; . 
\label{e10a}
\end{equation}
In contrast, one expects that at low densities and temperature, 
chiral symmetry is broken in the color
singlet channel, $m_\psi^2 < 0$; if all current
quark masses vanishes, ${\cal M} = 0$, the pattern is
\begin{equation}
\langle \psi_{a_L,b_R} \rangle \; = \; \psi_0 \, \delta_{a_L,b_R}
\; , 
\label{e11}
\end{equation}
as is consistent with  (\ref{e4b}).  
Chiral symmetry is restored 
at high density or temperature, $m_\psi^2 > 0$.  Ignoring the
coupling to other fields, $\psi$ still develops an expectation
value from the mass term, $\cal M$,
\begin{equation}
\langle \psi_{a_L,b_R} \rangle \; = \; \psi' \; m_a \; \delta_{a_L,b_R} \; ,
\label{e12}
\end{equation}

For color superconductivity, I assume that 
the color anti-triplet channel is favored
at high density \cite{son,sch,gap,rock,qcd,q3a,q3b,dirk}, with $m_\phi^2 < 0$,
and disfavored at low density, with $m_\phi^2 > 0$.  
How the chiral transition and color superconductivity are
coupled is one of the principle questions to be addressed.

I start with the case of two flavors.
For flavor $SU(2)$, 
${\bf 2}\times {\bf 2} = {\bf 1}_a + {\bf 3}_s$.
The color anti-triplet superconducting
field is then a flavor singlet \cite{bl}:
\begin{equation}
\phi_{i,R} \; = \;
\epsilon^{i j k} \; \epsilon^{a_R b_R} \;
\Phi_{j,k;a_R, b_R}  \; .
\label{e13}
\end{equation}
For two flavors I ignore 
the adjoint chiral field and the color sextet field for
color superconductivity, since they always
vanish: $\langle \widetilde{\psi} \rangle =
\langle \chi \rangle = 0$ at all densities.  
Under the abelian flavor symmetries, $\phi_{i,R}$ transforms
like $\Phi_R$, (\ref{e5}), {\it etc.}

Many interaction terms need to be added to ${\cal L}^0$; those which
violate $U_A(1)$ are especially interesting.  
For $N_f$ flavors of massless quarks,
the zero modes of an instanton with topological charge $Q$
generate an interaction between
$Q N_f$ right-handed quarks and $Q N_f$ left-handed anti-quarks \cite{hooft}.
From (\ref{e5}), the corresponding operators transform as
$\exp(2 i Q N_f \theta_A)$ under $U_A(1)$ rotations.
In vacuum instanton effects are large, since they give
the $\eta'$ its mass; thus they must continue to be important
in a hadronic phase, at small chemical potential.
Conversely, semiclassical methods are valid 
at large chemical potential, and it is certain then
that instantons are very dilute.  
At intermediate chemical potential, it is not clear 
how the density of instantons is correlated with
chiral symmetry breaking and color superconductivity.
I discuss what might happen if the density of instantons
is large in a chirally symmetric, color superconducting phase,
but this might not occur in QCD: the density of instantons
might drop precipitously when chiral symmetry is restored.

For two flavors, single instantons generate a determinental term
for the chiral fields \cite{hooft}, 
\begin{equation}
{\cal L}^I_\psi \sim - \, \det(\psi) \; ,
\label{e14}
\end{equation}
which is quadratic in the $\psi$'s. 
The superscript $I$ is used to denote
that the term is induced by instantons.
The overall minus sign in
(\ref{e14}) is important \cite{hooft}.
At $\theta = 0$, the instanton term not only acts to make 
the $\eta$ meson, which has 
spin-parity $J^P = 0^-$, massive, but it also drives
chiral symmetry breaking in the $0^+$ channel.

Single instantons generate a similar term
for the $\phi$'s \cite{twoa}, 
\begin{equation}
{\cal L}_{\phi}^I
\; \sim \;
- \, (\phi_{i,L}^* \, \phi^{}_{i,R} + \phi_{i,R}^* \, \phi^{}_{i,L}) \; .
\label{e15}
\end{equation}
As for ${\cal L}^I_\psi$, I write 
${\cal L}_{\phi}^I$ with an overall
minus sign, so that it acts to drive color superconductivity \cite{twoa}.

Besides the terms induced by single instantons, 
${\cal L}^I_\psi$ and ${\cal L}_{\phi}^I$, there are also
terms induced by instantons with topological charge two.
Operators induced by $Q=2$ instantons include
$({\cal L}^I_\psi)^2$, ${\cal L}^I_\psi {\cal L}_{\phi}^I$, and
$({\cal L}_{\phi}^I)^2$.  For two flavors these operators are
a curiosity, but they arise naturally for three flavors.

At high densities, where $U_A(1)$ is essentially restored,
one can write terms which respect axial $U_A(1)$ by squaring
each term in (\ref{e14}) and (\ref{e15}), such as
\begin{equation}
|\det(\psi)|^2 \;\;\; , \;\;\;
| \phi_{i,L}^* \, \phi^{}_{i,R}|^2 \; .
\label{e16}
\end{equation}
There are also terms which couple $\phi$ to $\psi$, such as
\begin{equation}
{\rm tr}
(\psi^\dagger \psi)\; \left( |\phi_R|^2 \; + \; |\phi_L|^2  \right) \; .
\label{e18}
\end{equation}
and 
\begin{equation}
\det(\psi)^* \; \phi_{i,L}^* \, \phi^{}_{i,R} \; + \; c.c. \; .
\label{e17}
\end{equation}
The effect of (\ref{e17}) and (\ref{e18}) is
to couple the transition for color superconductivity to that
for chiral symmetry breaking.  
Model dependent analyses indicate that
chiral symmetry breaking and color superconductivity can coexist 
for some range of densities \cite{twoa,twob,rssv}.
However, all such models manifestly leave out confinement: while
two quarks may like to bind together in a color superconducting
condensate, in a phase with chiral symmetry breaking --- and so
confinement --- this could well be overwhelmed by the
tendency of three quarks to form a color singlet baryon;
see, also, \cite{birse}.
On this basis, I assume that the sign of the coupling
constants in (\ref{e17}) and (\ref{e18}) is positive,
so that chiral symmetry breaking suppresses color superconductivity.

When color superconductivity occurs, and ${\cal L}_{\phi}^I$ is important,
the preferred condensate is
\begin{equation}
\langle \phi_{i,(R,L)} \rangle \; = \; {\rm e}^{i \theta_{R,L}} 
\; \phi_0 \, \delta_{i 3} \;\;\; , \;\;\;
\theta_R \; = \; \theta_L \; ;
\label{e19}
\end{equation}
a global color rotation is done to align the condensate
in the color-$3$ direction.
This breaks $SU(3)_c \rightarrow SU(2)_c$, and leaves flavor
unbroken; $\phi_0$ is real.
There are two types of correlations in these expectation values.
First, the phases of $\phi_{i,R}$ and $\phi_{i,L}$ are equal,
$\theta_R = \theta_L$.
Parity switches right and left handed fields, so if both
fields have the same phase, it implies that the condensate
has spin-parity $J^P=0^+$.
Secondly, with (\ref{e19}) the direction of the right
and left handed condensates are the same in color space \cite{sonst,ehhs}.  

What happens at high densities, when instantons are very dilute?
There is always some density of instantons about, and they
generate a term such as ${\cal L}_{\phi}^I$,
albeit with a small coefficient.  In this limit, $U_A(1)$
symmetry is effectively restored, and $\theta_R$
and $\theta_L$ are not correlated, except 
over very large scales.  This is the (approximate) spontaneous
breaking of parity \cite{bl,rg,cfl,prl,rssv,ehhs}.
Phrased in another way, the $\eta$ meson is very light:
its mass is determined by (\ref{e17}),
$m^2_\eta \sim m_u m_d$ \cite{prl}.

What about the coupling between the directions of 
$\phi_{i,R}$ and $\phi_{i,L}$ in color space?  
While the instanton term is no longer important,
there are many other terms in the effective lagrangian
which couple the color direction of the two condensates.  One example is
(\ref{e16}); in weak coupling, this first
appears at $\sim g^4$ \cite{sonst,ehhs}, where $g$ is the QCD
coupling constant.
Thus while the phase of the right and
left handed condensates are not (strongly) correlated at high density,
they are correlated in color; for a dynamical explanation, see \cite{ehhs}.

This completes my discussion for two flavors.  
In QCD, the case of interest for dense quark matter is really that of
three flavors.  
A chemical potential doesn't
matter until it is greater than the mass of a particle, so
there is no Fermi sea until the quark chemical potential $\mu$
is greater than one third of the nucleon mass, $\mu > 313 MeV$
(because of binding in nuclear matter, it is actually a little less).
As $\mu$ is always at least three times the strange quark mass,
any complete analysis must include three flavors.
Of course this counting is only valid in a chirally symmetric phase; with
chiral symmetry breaking, the constituent quark masses are large,
$\sim 313 MeV$, and strange baryons are suppressed.  In this
region my caveats about confinement apply.

For three flavors, the color anti-triplet superconducting field is
a flavor anti-triplet; for right handed particles, this is
\begin{equation}
\phi_{i,a_R} \; = \;
\epsilon^{i j k} \; \epsilon^{a_R b_R c_R} \;
\Phi_{j,k;b_R, c_R}  \; .
\label{e20}
\end{equation}
I also introduce the color sextet, flavor sextet superconducting field by
symmetrizing with respect to the color and flavor indices;
for right handed particles, 
\begin{equation}
\chi_{i,j;a_R,b_R} \; = \; 
\left(
(\Phi_{i,j;a_R,b_R} + 
(i \leftrightarrow j ) ) + (a_R \leftrightarrow b_R ) 
\right) \; .
\label{e21}
\end{equation}
My notation is somewhat confusing: for either color or flavor,
the indices on $\phi$ are
anti-triplet, while those on $\chi$ are triplet.

There are several terms which are special to three flavors.
For three flavors I keep track of all fields, including those
which are not favored to condense: the color
adjoint chiral field $\widetilde{\psi}$ and
the color sextet superconducting field $\chi$.
As will be seen, because of cubic operators they develop
expectation values when color superconductivity occurs.

I first consider operators induced by single instantons.
The simplest is a determinent
for the chiral fields, ${\cal L}_\psi^I$ in
(\ref{e14}).  This is just like that for two flavors,
except now it is cubic in the component fields $\psi_{a_L,b_R}$.
Analogously, there is also a determinental operator for 
three color adjoint chiral fields,
$$
{\cal L}_{\widetilde{\psi}}^I \; \sim \;
{\rm tr}\left(\det(\widetilde{\psi})\right)
$$
\begin{equation}
 \; \sim \;
d^{A B C} \; \epsilon^{a_R b_R c_R} \;
\epsilon^{a_L b_L c_L} \;
\widetilde{\psi}^A_{a_R a_L} \; 
\widetilde{\psi}^B_{b_R b_L} \; 
\widetilde{\psi}^C_{c_R c_L} \;  ,
\label{e21a}
\end{equation}
($d^{A B C}$ is the symmetric structure constant
for $SU(3)_c$) and between two color adjoint chiral fields and one
color singlet chiral field,
\begin{equation}
{\cal L}_{\psi \widetilde{\psi}}^I \; \sim \;
\epsilon^{a_R b_R c_R} \;
\epsilon^{a_L b_L c_L} \;
\psi_{a_R a_L} \; 
\widetilde{\psi}^C_{b_R b_L} \; 
\widetilde{\psi}^C_{c_R c_L} \;  .
\label{e21b}
\end{equation}

For the $\phi$ fields, in obvious analogy there are
two cubic operators which are invariant under the nonabelian symmetries,
but transform under $U_A(1)$:
\begin{equation}
H_R \; = \; \det(\phi_{i,a_R}) \;\;\; , \;\;\;
H_L \; = \; \det(\phi_{i,a_L}) \; ,
\label{e22}
\end{equation}
For later reference, I introduce
\begin{equation}
H_\pm \; = \; \frac{1}{2} \left( H_R \pm H_L \right) \; .
\label{e23}
\end{equation}
$H_\pm$ has spin-parity $J^P = 0^\pm$, so $H_+$ has the quantum numbers of the
$H$-dibaryon \cite{jaffe}.  However, unlike $\det(\psi)$
and $\det(\widetilde{\psi})$, 
$H_R$ and $H_L$ cannot appear in an effective lagrangian, because
they transform not only under the anomalous $U_A(1)$,
but also under the (good) $U(1)$ symmetry for quark number \cite{con}.
Using only the $\phi$ fields, one can construct terms which are
invariant under all symmetries except $U_A(1)$:
\begin{equation}
{\cal L}^I_H \; \sim \; H_L^* \, H_R \; + \; H_R^* \, H_L \; .
\label{e24}
\end{equation}
This is like ${\cal L}_{\phi}^I$ 
for two flavors, but given the transformation properties of 
$\phi$ under $U_A(1)$, (\ref{e5}), ${\cal L}^I_H$
is not induced by single instantons, but  
by instantons with topological charge two.

An operator induced by a single instanton is given
by combining two $\phi$'s and one 
color singlet chiral field, $\psi$ \cite{cfl,q3a}:
\begin{equation}
{\cal L}_{\phi \psi}^I \; \sim \; - \left(
\phi^*_{i,a_L} \; \phi^{}_{i,b_R} \; \psi^{}_{a_L,b_R}\; + \; {\rm c.c.}
\right) \; .
\label{e25}
\end{equation}
I assume the sign is negative, as it is for two flavors.
Single instantons also induce a similar term
between two $\phi$'s and one color adjoint chiral field, 
$\widetilde{\psi}$,
\begin{equation}
{\cal L}_{\phi \widetilde{\psi}}^I \; \sim \; 
\phi^*_{i,a_L} \; t^C_{i,j} \; \phi^{}_{j,b_R} \;  
\widetilde{\psi}^C_{a_L,b_R} \; + \; {\rm c.c.} \; .
\label{e25a}
\end{equation}
I do not know the sign of (\ref{e25a}), but it is unimportant.  

As for two flavors, when chiral symmetry is broken
in the color singlet channel, (\ref{e11}),
${\cal L}_{\phi \psi}^I$ helps to generate
color superconductivity.  I assume that quartic
terms in the potential, such as
\begin{equation}
{\rm tr}(\psi^\dagger \psi) \; 
{\rm tr}\left(\phi_R^\dagger \phi^{}_R 
\; + \; \phi_L^\dagger \phi^{}_L \right) \; ;
\label{e26}
\end{equation}
where $tr (\phi_R^\dagger \phi^{}_R) =
\phi_{i,a_R}^* \phi_{i,a_R}$, {\it etc.},
are sufficiently large and of positive sign, so that phases
with chiral symmetry breaking and color superconductivity do not
overlap.

This terminology is imprecise.  Consider what happens when chiral
symmetry is restored, $m_\psi^2 > 0$, so the expectation value of
$\psi$ is naively that of (\ref{e12}).  
The preferred condensate is color-flavor locked \cite{cfl}:
\begin{equation}
\langle \phi_{i,a_{R,L}}\rangle 
\; = \; {\rm e}^{i \theta_{R,L}} \;
\phi_0 \, \delta_{i,a_{R,L}} \;\;\; , \;\;\;
\theta_R \; = \; \theta_L \; ;
\label{e27}
\end{equation}
global color and flavor rotations are done to
make the condensates diagonal.
This patterns breaks $SU(3)_c \times SU(3)_L \times
SU(3)_R \times U_A(1) \times U(1) \rightarrow SU(3)$.
Because of ${\cal L}_{\phi \psi}^I$, 
the right and left handed condensates have
the same phase, so the condensate has $J^P=0^+$.

I remark that although the physics is very different,
formally the pattern of symmetry breaking for color-flavor
locking in (\ref{e27}) is identical to that for chiral symmetry breaking
in (\ref{e11}) \cite{proc}.  As for chiral
symmetry breaking, (\ref{e27}) is not the only possible way in
which color superconductivity could occur; for example,
one might have 
$\langle \phi_{i,a_R}\rangle = \phi_0 \delta_{i,3}\delta_{i,a_R}$
\cite{prl}.  It is easy to argue that this is disfavored \cite{cfl,prl}:
such a condensate leaves at least 
two different colors and flavors ungapped, while with color-flavor locking,
all colors and flavors of quarks are gapped.  

Because of the instanton terms, however,
when color superconductivity occurs, 
${\cal L}_{\phi \psi}^I$ and 
${\cal L}_{\phi \widetilde{\psi}}^I$
become terms which are {\it linear} in $\psi$ and $\widetilde{\psi}$,
respectively.  
Consequently, expectation values
for $\psi$ and $\widetilde{\psi}$ 
are automatically generated when $\phi$ condenses.
For the color singlet chiral
field, this means that the expectation
value of $\psi$ never vanishes, even at high density
in the chiral limit, ${\cal M} = 0$.   The color adjoint
chiral field also develops an expectation value, as is seen
in a three flavor instanton model \cite{rssv}.

Since the expectation value of
$\psi$ is always nonzero, there is no gauge invariant order
parameter which distinguishes a phase
with chiral symmetry breaking from one with color superconductivity,
and at least formally, there is a continuity
between strange hadronic matter and strange quark matter \cite{con}.  
One might wonder if $\det(\widetilde{\psi})$,
(\ref{e21a}), provides such an order parameter, but even though
$\langle \widetilde{\psi}\rangle = 0$ in the phase with
chiral symmetry breaking, assuming
the quark expectation value of (\ref{e4b}),
$\langle \det(\widetilde{\psi})\rangle \sim \psi_0^3 \neq 0$.

Even so, I argue that in QCD, phases with chiral symmetry
breaking and color superconductivity appear to be rather different.
In a hadronic phase, the relationship between chiral symmetry
breaking and confinement helps us to understand the central
mystery of nuclear physics: why the
nuclear binding energy, $\sim 16 MeV$, is so small relative
to {\it any} other scale in QCD \cite{weinberg}. The scale of hadronic
superfluidity is smaller still, $\leq 3 MeV$ \cite{super}.
If, as originally believed \cite{bl},
the gaps for color superconductivity are also $\sim 1 MeV$,
then continuity between hadronic and quark matter 
is automatic.  From recent work
with effective models, however, it appears
that the color superconducting gaps
are natural on a QCD scale, $\sim 100 MeV$
\cite{gen,cfl,con,twoa,twob,abr,rssv}, and so huge
relative to the hadronic gaps.  If true, I assume that this 
disparity in scales, by almost two orders of magnitude, 
is due to confinement.  

Thus I distinguish between a phase driven by chiral symmetry breaking,
where $m_\psi^2 < 0$ and $m_\phi^2 > 0$, from a phase driven
by color superconductivity, with $m_\psi^2 > 0$ and $m_\phi^2 < 0$.
At high density, the instanton terms
${\cal L}_{\phi \psi}^I$ and 
${\cal L}_{\phi \widetilde{\psi}}^I$ are
very small, so the expectation values of $\psi$ and
$\widetilde{\psi}$ are negligible.

At intermediate densities,
the instanton term ${\cal L}_{\phi \psi}^I$
has several interesting effects.
Remember that the $\phi$ field is anti-triplet in the flavor indices.  Thus
the strange component,$\phi_{i,3}$ is an up-down condensate, while
the up and down components,
$\phi_{i,1}$ and $\phi_{i,2}$, are condensates of down-strange and
up-strange, respectively.  Since $m_s \gg m_u , m_d$, 
${\cal L}_{\phi \psi}^I$ 
is greatest for the up-down condensate,
$\sim - m_s \phi^*_{i,3} \phi_{i,3}$, and smallest for the
down-strange and up-strange condensates,
$\sim - m_u \phi^*_{i,1} \phi_{i,1}$ and 
$\sim - m_d \phi^*_{i,2} \phi_{i,2}$.
This is reasonable: because 
of the difference in the quark masses, it is easiest for
color superconductivity to occur between up and down quarks, and hardest
for it to from between up or down and strange quarks.  With the
overall minus sign, this is exactly what 
${\cal L}_{\phi \psi}^I$  does.

As for ${\cal L}_{ \phi}^I$ with two flavors,
for three flavors ${\cal L}_{\phi \psi}^I$ correlates the overall
phases of the right and left handed condensates.  
Because of the difference in the quark masses, though,
it is most effective for the up-down condensate, and least effective
for the up-strange and down-strange
condensates, by a factor of $m_s/m_{u,d} \sim 20$.
This implies that in a phase driven by color superconductivity,
{\it if} instantons are important, then 
the (approximate) spontaneous violation of parity is {\it smallest}
for the up-down condensate, and {\it greatest} for the 
up-strange and down-strange condensates.
If instantons are not important, then
all three condensates exhibit the same (approximate)
parity violation, and the $\eta'$ is the lightest 
pseudo-Goldstone boson \cite{cas,sonst}.

It is not clear how to observe the (approximate) spontaneous
violation of parity in the up-strange and down-strange condensates.
As an effect from a Fermi sea, this appears most directly in 
baryons; the pattern above suggests effects are large for
$\Lambda$ baryons, and negligible for any baryons which have
two quarks of the same flavor.  Any effect
is obscured by the fact that even 
in vacuum, the decays of the $\Lambda$ 
are not parity conserving, for reasons which are not well understood.

What about effects in a phase with (approximate) $U_A(1)$
symmetry?  The color directions of the right and left
handed fields are correlated through terms of quartic order
in the potential, including
\begin{equation}
( \phi^*_{i,a_R} \, \phi^{}_{i,b_L} ) \; 
( \phi^*_{j,b_L} \, \phi^{}_{j,a_R} ) \; ,
\label{e28}
\end{equation}
which is analogous to the quartic coupling for two flavors in
(\ref{e16}).  Mass dependence for the $\phi$'s enter through
terms such as (\ref{e26}) and 
\begin{equation}
|\phi^*_{i,a_L} \; \phi^{}_{i,b_R} \; \psi^{}_{a_L,b_R} |^2 \; .
\label{e29}
\end{equation}
These terms are analogous to those using 
nonlinear effective lagrangians \cite{cas,sonst}.

For three flavors, condensation of the color anti-triplet
superconducting field $\phi$ also drives
that of the color sextet superconducting
field $\chi$.  Consider the operators \cite{prl,proc}
\begin{equation}
{\cal L}_{\phi \chi} 
\; \sim \; 
H_R^* \; \phi^{}_{i,a_R} \; \phi^{}_{j,b_R} \; \chi^{}_{i,j;a_R,b_R} \; 
+ \; {\rm c.c.} \; .
\label{e30}
\end{equation}
and
\begin{equation}
{\cal L}_{\phi \chi}^I
\; \sim \; 
H_L^* \; \phi^{}_{i,a_R} \; \phi^{}_{j,b_R} \; \chi^{}_{i,j;a_R,b_R} \; 
+ \; {\rm c.c.} \; .
\label{e31}
\end{equation}
Both operators are invariant under the $U(1)$ of quark number;
${\cal L}_{\phi \chi}$ is invariant under $U_A(1)$,
while ${\cal L}_{\phi \chi}^I$ is induced by instantons with
topological charge two.
When $\phi$ condenses according to (\ref{e27}), 
${\cal L}_{\phi \chi}$ and ${\cal L}_{\phi \chi}^I$
become terms {\it linear} in $\chi$, and generate
an expectation value for $\chi$.  
The terms in (\ref{e30})
and (\ref{e31}) explain why for three flavors
the preferred condensate always contains some (small) piece
in the repulsive, color sextet channel \cite{cfl,q3a,q3b,abr,rssv}.  

Since ${\cal L}_{\phi \chi}$ is invariant under $U_A(1)$, it is
present even at high density.  This is in contrast to
the instanton induced operators
${\cal L}_{\phi \chi}^I$, ${\cal L}_{\phi \psi}^I$ and 
${\cal L}_{\phi \widetilde{\psi}}^I$, 
which are small at high density.
There is no anomaly in perturbation theory, so in weak coupling
one finds that because of ${\cal L}_{\phi \chi}$,
$\chi$ condenses, but not
$\psi$ or $\widetilde{\psi}$ \cite{q3a,q3b}.

I conclude this section
by discussing the fields related to $H$-superfluidity,
$H_+$ and $H_-$ \cite{con,prl}.  
Each is a complex valued scalar field with two real degrees
of freedom.  In a color superconducting phase, one mode
of $H_+$ is the Goldstone boson for the spontaneous breaking
of the $U(1)$ symmetry of quark number, and so is massless.
The other $H_+$ mode is massive except near a second order
transition where $\langle H_+ \rangle \rightarrow 0$, at which point
both modes form a $O(2)$ multiplet.  The $H_-$ field is
like $H_+$, except that both components obtain a mass
from instantons, from ${\cal L}^I_H$ in (\ref{e24}).

Since $H_+$ only cares about the (spontaneous) breaking of the
$U(1)$ for quark number, it is
{\it not} affected by nonzero, nondegenerate quark masses.
All that matters is that all three colors and flavors of quarks
become color superconducting.
Alternately, one may consider separate fields for the three
condensates, and construct 
the operator analogous to $H_+$; see [21] of \cite{prl}.

It is possible to construct gauge invariant, superfluid order parameters for
two flavors by using both the particle,
$\phi^+_{i,R}$, and anti-particle, $\phi^-_{i,R}$, condensates:
\begin{equation}
\epsilon^{i j k} \; \phi^+_{i,L} \; \phi^+_{j,R} \; \phi^-_{k,L} \; .
\label{e32}
\end{equation}
Like $H_+$ and $H_-$, this is invariant under all but the abelian
flavor symmetries.  However, there is no reason to believe that this
quantity is ever nonzero, since all three fields most likely
lie in the same direction in color space.  Thus $H$-superfluidity
is uniquely a consequence of color superconductivity
through color-flavor locking for $2+1$ flavors.

\section{Phase Diagrams}

In this section I begin by proposing phase diagrams for the effective
three dimensional theories which describe the phase 
transition to color superconductivity at nonzero temperature.  
I then conjecture how this might relate to the phase diagram
in the plane of chemical potential and temperature.

I assume that the density of instantons is always large, so that
the right and left handed condensates are equal.  Properly,
I should allow the right and left handed condensates to differ,
but even at high density, this does not appear to 
affect the order of the phase transition.
The moral of the preceeding section is that for three flavors,
because of cubic terms involving two $\phi$'s and 
the other fields --- either $\psi$,
$\widetilde{\psi}$, or $\chi$ --- expectation values
these other fields are generated by color superconductivity.
Even so, assuming that these other fields all have positive
mass squared, then as
the $\phi$ field becomes critical, $m_\phi^2 \rightarrow 0$,
all of the other fields remain noncritical.  Thus we can
safely neglect all fields except for the color anti-triplet
superconducting field.

For two flavors, I denote the condensate field as
$\overline{\phi}_i \equiv \overline{\phi}_{i,R} = \overline{\phi}_{i,L}$;
the effective lagrangian is
\begin{equation}
{\cal L}_2 \; = \;
\frac{1}{2} \; {\rm tr}\left( G_{\mu \nu}^2 \right) 
\; + \; |D_\mu \overline{\phi}|^2 \; + \;
m_\phi^2 \; |\overline{\phi}|^2 \; + \;
\overline{\lambda} \; \left( |\overline{\phi}|^2 \right)^2 \; ;
\label{e33}
\end{equation}
$G_{\mu \nu}$ is the field strength for the gauge field $A_\mu$,
and  $D_\mu = \partial_\mu + i g A_\mu$ is the covariant derivative
for an anti-triplet color field $\overline{\phi}_i$.  I distinguish
the condensate field $\overline{\phi}_i$ from $\phi_i$ in the 
previous section, due to an overall rescaling explained below, 
(\ref{e37}).
I need the coupling constants for the effective three dimensional
theory near the transition temperature, but I assume that these
are just the temperature $T$ times that those in four dimensions,
which is approximately true.

The phase transition occurs as $m_\phi^2 \rightarrow 0$; in this
case it is natural to introduce the ratio of the
$\overline{\phi}$ coupling constant, $\overline{\lambda}$,
to that for the gauge field, $g$,
\begin{equation}
\lambda \; = \; \frac{\overline{\lambda}}{g^2}.
\label{e34}
\end{equation}
This ratio has a more physical interpretation.  At zero temperature,
where $m_\phi^2 < 0$, the expectation value of $\overline{\phi}$ is
$\overline{\phi}_0 \sim (-m^2_\phi/\overline{\lambda})^{1/2}$.
In the broken phase, the 
Higgs mass is the mass for the $\overline{\phi}$ field,
$\sim m_\phi$, while the gluon has a mass
$m_A \sim g \overline{\phi}_0$.  Thus
$\lambda$ is proportional to the Higgs mass divided by the gluon
mass, squared: $\lambda \sim m_H^2/m_A^2$.

This is in contrast to what happens at zero temperature,
where $m_\phi^2$ is tuned to vanish by hand \cite{coleman,amit,arnoldetal}.
For $g=0$, the $\beta$-function for
$\overline{\lambda}$ has an infrared stable fixed point at the origin.  
When $g \neq 0$, however, $\overline{\lambda}$ cannot flow into the origin;
instead, it flows from positive to negative values, 
which then generates a first order transition.
Naively, one expects that $\lambda$
is a free parameter, but because of dimensional transmutation,
this is an illusion.  By letting the coupling
constants flow, one can always go from a regime with large
$\lambda$ to one with small $\lambda$.  Physically, the
square of the ratio of the masses for the Higgs to the gluon fields is
not a free parameter, but is fixed, $\sim g^2$.

There is no dimensional transmutation in three dimensions, so 
$\lambda$ is a free parameter: different values of $\lambda$
correspond to different theories at zero temperature.
We can then consider the phase diagram
as a function of $\lambda$ \cite{hlm,other,ginsparg,march,yaffe,philip}.
For small $\lambda$, 
fluctuations in the gauge field dominate,
and a one loop analysis reliably indicates a 
first order phase transition.
This is the ``type-I'' regime of ordinary superconductivity.
As $\lambda$ increases, one moves into the 
``type-II'' regime, where fluctuations in the scalar field become
important.  An expansion from $4-\epsilon$ dimensions to three dimensions
predicts that for large $\lambda$, that the transition
is driven first-order by fluctuations in the scalar fields.
Consequently, the simplest hypothesis is an unbroken line of first
order transitions for all $\lambda$.

This is not what lattice simulations find \cite{ew}.  There is
indeed a first order transition for small $\lambda$, but
as $\lambda$ increases, the strength of the first order
transition decreases, until it ends at a critical point,
at $\lambda_c$.  
The critical point occurs when the masses of the gauge and scalar fields are
approximately equal.  For $\lambda> \lambda_c$,
there is no first order transition,
only a smooth crossover between the two phases.

The existence of a critical end point is certainly possible.
Since $\overline{\phi}$ is in the fundamental representation of the gauge
group, there is no gauge invariant order parameter which 
distinguishes between the two phases \cite{compl}.
The question, however, is why in the type-II regime
does the first order transition at small $\epsilon$ 
turn into a smooth crossover at $\epsilon = 1$?

One possibility is simply that the $\epsilon$-expansion breaks down
at large $\epsilon \sim 1$.
Nevertheless, I {\it assume} that the $\epsilon$-expansion is
reliable in its prediction of a fluctuation induced transition
when the theory only involves scalar fields.
Numerous examples are known in condensed matter physics \cite{other};
for careful analyses in models 
where the strength of the first order transition can be
controlled, see \cite{yaffe}.

Instead, I suggest that 
the $\epsilon$-expansion fails uniquely for theories of
scalars coupled to gauge fields.
Since the theory is three-dimensional, when
the vacuum expecation value of the scalar field vanishes, one
inevitably enters a strongly coupled phase of the theory.  In
this phase, the proper way to think of the spectrum is in terms
of gauge invariant excitations, such as glueballs and mesons
formed from scalars \cite{philip}.  A fluctuation induced first
order transition occurs when
the quartic couplings for the scalar run from positive to negative values.
For this to happen, however, the couplings must flow.
Perhaps the crossover regime is simply a manifestation
of confinement in three dimensions: when $m_\phi^2 \rightarrow 0$,
scalars become strongly bound into relatively heavy mesons.
If the scalars are heavy, the scalar self couplings never run by
much, even as $m_\phi^2 \rightarrow 0$.
That is, confinement in three dimensions
``eats'' the running of the coupling constants.  
While only a qualitative explanation, it is reasonable
that crossover begins when the (zero temperature) masses
of the Higgs and gauge fields are approximately equal.  
How (three-dimensional)
confinement can stop the running of (effective) coupling
constants is analogous to how, in four dimensions, the strong coupling
constant might freeze at low momenta \cite{freeze}.

The complete phase diagram can then be sketched.
Consider the limit of large $\lambda$; this may not
make sense in the continuum (because of triviality bounds), 
but is perfectly reasonable on the lattice.
For infinite $\lambda$, the gauge fields decouple, and
there is only a scalar field; the universality class
is that of an $U(3)$ vector, which is the same as 
an $O(6)$ vector.  Thus there is
a second order phase transition when $\lambda = \infty$.
At large but finite value of $\lambda$, however, 
confinement presumably eats the running of the 
scalar coupling constants, so the theory exhibits crossover for 
$\lambda_c < \lambda < \infty$.  
This phase diagram, from \cite{arnoldetal}, is
illustrated again in fig. (1): there are critical points at
$A_2$, where $\lambda = \lambda_c$, and at $B_2$, where
$\lambda = \infty$.

\begin{figure}
\begin{center}
\epsfxsize=4cm
\epsfysize=4cm
\leavevmode
\hbox{ \epsffile{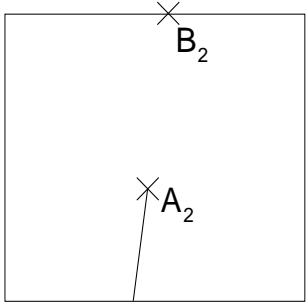}}
\end{center}
\caption{Phase diagram for one $\overline{\phi}_i$ field coupled to a $SU(3)$
gauge field.  In figs. 1 and 2, 
the y-axis is $\lambda$, from $0$ to $\infty$, 
and the x-axis is $\sim m_\phi^2$.}
\label{f1}
\end{figure}

Consider now the model for three flavors, where the anti-triplet
condensate field is $\overline{\phi}_{i,a}$:
$$
{\cal L}_3 \; = \;
\frac{1}{2} \; {\rm tr}\left( G_{\mu \nu}^2 \right) \; + \;
{\rm tr} \left(D_\mu 
\overline{\phi}^\dagger D_\mu \overline{\phi} \right) \; + \;
m_\phi^2 \; {\rm tr}\left( \overline{\phi}^\dagger \overline{\phi}\right)
$$
\begin{equation}
\; + \; \overline{\lambda}_1 \; 
\left( {\rm tr} 
\left( \overline{\phi}^\dagger \overline{\phi} \right) \right)^2 \;
+ \; \overline{\lambda}_2 \; 
{\rm tr} \left( \overline{\phi}^\dagger \overline{\phi} \right)^2 \; .
\label{e35}
\end{equation}
There are now two quartic coupling constants, 
$\overline{\lambda}_1$ and $\overline{\lambda}_2$,
so there are two $\lambda$ parameters like that of 
(\ref{e34}); for simplicity I speak only of one,
assuming that $\overline{\lambda}_2 \neq 0$, so that there is not
an accidental $O(18)$ symmetry.
At $\lambda = \infty$, gauge fields can be neglected,
and $\overline{\phi}_{i,a}$ is a 
$SU(3) \times SU(3) \times U(1)$ vector field.
In $4-\epsilon$ dimensions, this has a fluctuation
induced first order transition \cite{susu}.
Assuming this persists to three dimensions, 
there is a first order transition at $\lambda = \infty$;
even with confinement at $\lambda < \infty$, 
the first order transition continues for
some finite range of large $\lambda$.  Assuming
a crossover regime for intermediate
$\lambda$, the phase diagram, illustrated in fig. (2),
has a critical end point for small $\lambda$, at a point
$A_3$, and for large $\lambda$, at a point $B_3$.  
Both critical end points are in the Ising universality class in
three dimensions,
as are points $A_2$, $B$, $A_3$, and $B_3$ in figs. (3) and (4).

\begin{figure}
\begin{center}
\epsfxsize=4cm
\epsfysize=4cm
\leavevmode
\hbox{ \epsffile{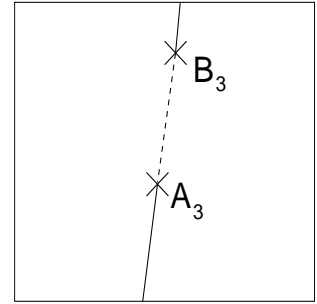}}
\end{center}
\caption{Phase diagram for 
$\overline{\phi}_{i,a}$ coupled to a $SU(3)$ gauge field.
}
\label{f2}
\end{figure}

Between $A_3$ and $B_3$, there is 
a gauge invariant order parameter, $H_+$, which distinguishes
between the two phases.  The expectation value of $H_+$ either
goes to zero continuously or discontinuously.  If the latter,
there is a first order transition, and no crossover regime.
Assuming there is a crossover regime for the nonabelian fields, 
between the critical end points $A_3$ and $B_3$ there must be
a {\it line} of second order phase transitions, at which
the $H_+$ field becomes critical.  This is in the universality class
of an $O(2)$ vector in three dimensions.

This critical line is analogous to that found
by Kajantie {\it et. al.} in their study of an adjoint scalar
field coupled to a $SU(3)$ gauge field \cite{adj}.  
(For the model of \cite{adj}, there is only one
quartic scalar coupling, and so the point $B_3$ lies at
$\lambda = \infty$.)  In the
present case, one might wonder why the critical fluctuations
for the $H_+$ are not eaten by confinement.  The same
comment applies to the adjoint model of \cite{adj}, and is
easy to dismiss.  For color superconductivity, the fluctuations
in $H_+$ are associated entirely with fluctuations in the $U(1)$
symmetry of quark number.  The nonabelian $SU(3)_c$ gluons cannot eat
the running of the coupling constant for 
the $U(1)$ of quark number because they cannot ``taste'' it;
the $H_+$ field is neutral under any $SU(3)_c$ transformation.  Thus
near the critical line for $H$-superfluidity,
the only critical modes are those for $H_+$.

These effective theories can be directly related to color
superconductivity, although there is one surprise.  
As is known in ordinary superconductivity \cite{bl}, the condensate
field does not have canonical normalization; the effective
lagrangian for $\phi$ is
\begin{equation}
{\cal L} \; = \;
\frac{\mu^2}{\phi_0^2} \; |D_\mu \phi|^2
\; - \; \mu^2 |\phi|^2 \; + \; 
\frac{\mu^2}{\phi_0^2} \; \left(|\phi|^2\right)^2 \; + \ldots
\label{e36}
\end{equation}
The terms in (\ref{e33}) are only correct up to coefficients
of order one, and I am sloppy about which quartic terms enter,
because all I really care about is how the scales in the
problem --- the chemical potential, $\mu$, and the value
of the condensate at zero temperature, $\phi_0$ --- enter.
What is interesting is that because particles have an 
energy $\sim \phi_0$ near the Fermi surface, 
the kinetic and quartic terms in the potential
have factors of $\mu^2/\phi_0^2$.  
This can be understood as follows.  The mass term, $\sim |\phi|^2$,
is not singular, with an overall mass dimension set by the chemical
potential, $\mu$.  Expanding the two point function of $\phi$ in
momentum, the natural scale for the momenta to vary is over $\phi_0$;
thus the kinetic term is $\mu^2 |\partial_\mu \phi|/\phi_0^2$.
The gauge invariant generalization is
$\mu^2 |D_\mu \phi|/\phi_0^2$, which includes a quartic
interaction between two $\phi$'s and two
$A_\mu$'s $\sim \mu^2/\phi_0^2$.  Thus it is not surprising
that the quartic interaction 
between four $\phi$'s also has an overall factor $\sim \mu^2/\phi_0^2$.

To relate this to the effective lagrangians in (\ref{e33}) and
(\ref{e35}), it is necessary to rescale the fields and coupling
constants, so that
\begin{equation}
\phi \;\sim \; \frac{\phi_0}{\mu} \; \overline{\phi} \;\;\; , \;\;\;
\lambda \; \sim \; \left(\frac{\phi_0}{g \mu}\right)^2 \; .
\label{e37}
\end{equation}
It is also clear from the form of the potential that terms
of higher order, such as a six point term,
$(\mu/\phi_0)^4 (|\phi|^2)^3
\rightarrow (\phi_0/\mu)^2 
(|\overline{\phi}|^2)^3$, are just as important as the
quartic term.  Indeed, all of the effective terms in the previous
section should be multiplied by corresponding powers of
$\mu/\phi_0$.  None of the results change qualitatively,
since they were a consequence of symmetry, and not of the
assumption of limiting oneself to operators with the smallest
mass dimension.

These powers of $\mu/\phi_0$ imply that, as in ordinary
superconductivity, generally a Landau-Ginzburg approach is
a terrible approximation.  The one exception is near a point
of phase transition, where $\langle \phi \rangle \rightarrow 0$.
In this case, fluctuations are controlled by the term with
the largest mass dimension; {\it i.e.}, cubic and quartic terms.

Using the conjectured results for the phase diagram of the
effective theories in three dimensions, we can then draw
cartoons for the possible phase diagrams of color superconductivity
in the $\mu-T$ plane.  At large $\mu$, where by asymptotic
freedom $g(\mu)$ is small, the gap is
exponentially small in $1/g$ 
\cite{prl,son,sch,gap,rock,qcd,cdw,q3a,q3b,ehhs,dirk}:
\begin{equation}
\phi_0 \; = \;
512 \; \pi^4 \; 
\left( \frac{2}{g^2 N_f}\right)^{5/2} \;
\exp\left(- \frac{3 \pi^2}{\sqrt{2} g} \right) \; \mu \; b_0' \;;
\label{e38}
\end{equation}
$b_0'$ is a pure number, determined in \cite{rock}.
Notice that the gap decreases as the number of (massless) flavors, $N_f$,
increases.  In mean field theory, the transition temperature
is as in the theory of Bardeen, Cooper, and Schrieffer,
$T_c \sim .567 \phi_0$, and is of second order \cite{gap}.
Thus at a large but fixed value of $\mu$, as the temperature increases,
there is a transition at which superconductivity for three flavors
evaporates, and then a higher temperature at which that for
two flavors evaporates.  When fluctuations are included, both transitions
turn first order.  Since the condensate is small, 
the effective coupling $\lambda \sim (\phi_0/g\mu)^2$ is very small,
and the theory is in a regime of extreme type-I,
with a tiny latent heat $\sim \lambda$.

As $\mu$ decreases, $\lambda$ increases, so one moves up the
phase diagrams of figs. (1) and (2).  The crucial question is
how color superconductivity matches onto the chiral phase transition.
I henceforth {\it assume} that the type-I regime ends before the
chiral phase transition.  In this case, there are critical end points
for color superconductivity, at points $A_2$ and $A_3$, respectively,
in figs. (3) and (4); they correspond 
precisely to the same points in figs. (1) and (2).  

Now consider the opposite limit, working up for small chemical
potential, $\mu$.  I assume that at zero temperature there
is a first order chiral transition, at a point $C$ in figs. (3) and
(4).  As the temperature increases, the chiral transition occurs
at smaller $\mu$, so the chiral transition bends back,
extending to a point
$E$, which is a critical end point for the chiral phase transition
\cite{rss}.  

How do the phase transitions for color superconductivity and
the chiral transition match onto each other?
Based upon the discussion in the previous
section, I assume that at zero temperature
the chiral transition coincides with that for color superconductivity.  
There are then two cases: if
the gap for color superconductivity at the point $C$ is small relative
to the (current) strange quark mass, then 
increasing $\mu$ at $T=0$, one first 
enters a phase in which only up and
down quarks superconduct \cite{abr}, and then
a phase in which all three flavors superconduct.
This is illustrated
in fig. (3).  Following \cite{prl}, at zero
temperature the transition from 
two flavor to three flavor superconductivity is of first order.  

(A crucial assumption in \cite{prl} is that hard dense
loops give the gluons a ``mass'', so they decouple from
the phase transition.  One might question if this remains
true in strong coupling; even with the hard dense loop mass,
perhaps the four dimensional gluons eat the running of the
coupling constants for the condensate field?  
With some effort, this can be analyzed on the lattice:
it would be necessary to add dummy fields to generate hard
dense loops for the gluons, and then couple the gluons
to a condensate field.)

\newpage
\begin{figure}
\begin{center}
\epsfxsize=5cm
\epsfysize=5cm
\leavevmode
\hbox{ \epsffile{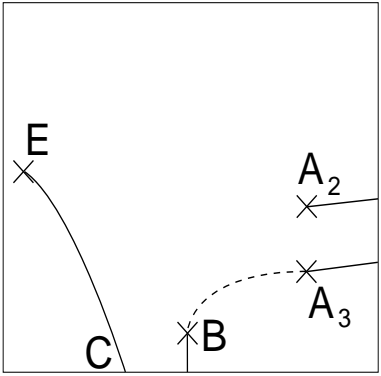}}
\end{center}
\caption{Cartoon of the QCD phase diagram in the small gap limit.
In figs. (3) and (4), the y-axis is temperature, and the x-axis
quark chemical potential.}
\label{f3}
\end{figure}

Alternately, the gap at the point $C$ could be large relative to
the strange quark mass.  In this limit,
one goes directly from a phase with (large) chiral symmetry
breaking, to one with three flavor color superconductivity,
fig. (4).  As discussed in the previous
section, there need not be a true phase transition at the
point $C$ \cite{con}; I assume there is, based on the disparity
in scales for superfluidity between hadronic and quark matter.

\begin{figure}
\begin{center}
\epsfxsize=5cm
\epsfysize=5cm
\leavevmode
\hbox{ \epsffile{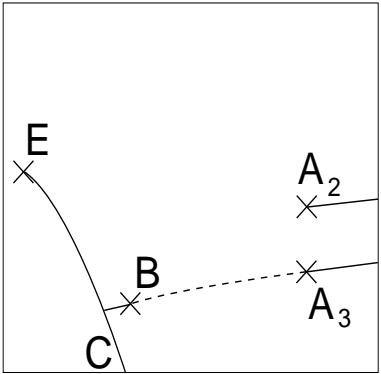}}
\end{center}
\caption{Cartoon of the QCD phase diagram in the large gap limit.}
\label{f4}
\end{figure}

The crucial assumption in figs. (3) and (4) is
that the color superconductor does not remain in the type-I
regime, but enters the type-II, or crossover, regime.
Thus the line for two flavor superconductivity terminates in
a critical endpoint, $A_2$, with no true phase transition
for $\mu$'s less than some value.  
For three flavor superconductivity, if one enters the crossover
regime there must be a critical line
for $H$-superfluidity.  Even if one enters the type-II regime,
one may not reach the second critical endpoint, $B_3$ of
fig. (2).  In figs. (3) and (4) I assume not.  In fig. (3),
the point $B$ represents where the critical theory goes from
being four dimensional to a three dimensional theory in the crossover
regime.  In fig. (4), since a critical line probably does not
attach directly to the line of first order chiral transitions,
the point $B$ represents the end of a first order line induced by the
chiral transition.  

There is a caveat to the phase diagrams of figs. (3) and (4).
Even in a confined phase, 
as long as strange quarks populate a Fermi sea,
$\Lambda$ baryons may well be superfluid, so that at $T=0$, 
$\langle H_+ \rangle \sim \langle \Lambda \Lambda \rangle\neq 0$.  
The phase
transition for such ``hadronic'' $H$-superfluidity is probably of
second order for all $\mu$ and $T$.  As discussed
in sec. II, though, hadronic superfluid gaps
appear to be much smaller than color superconducting gaps.
Since critical temperatures are proportional to the gap,
in the $\mu-T$ plane the lines for hadronic $H$-superfluidity
lie very close to the zero temperature axis.
Similarly, even between the two first order transitions
at $T=0$ in fig. (3), 
$\langle H_+ \rangle \neq 0$.
This is a phase in which only up and down,
but not strange, quarks, superconduct.  In this region, I also
assume that $\langle H_+ \rangle$ is small, on the order of that in
the hadronic phase.

I stress that I assume that the theory goes from the type-I
to the type-II regime as $\mu$ decreases.  It is conceivable that the 
color superconducting transitions remain
in the type-I regime for all $\mu$.  In this instance, the point
$A_2$ would reach all the way to the chiral line, and 
$A_3$ and $B$ would coincide, with no critical line for
$H$-superfluidity.   Alternately, it is also possible that
at small $\mu$ the theory goes so deep into the type-II regime
that the point $B$ in figs.
(3) and (4) coincides with the critical end point, $B_3$, in fig. (2).  

My arguments are admittedly speculative,
and meant only to suggest what the QCD phase diagram might look like.  
While at present the lattice cannot tell us 
about QCD with $\mu \neq 0$, it can study the effective
theories of relevance to color superconductivity.  Moreover,
by using perturbation theory in QCD, one can work down from
large $\mu$ to match onto the lattice results.
How far this can be pushed at small $\mu$ is open to question,
but is testable.  The most interesting question is where
the critical endpoint $A_3$ is, since that tells us if and when
the critical line for $H$-superfluidity begins.

{\it Acknowledgements}: I thank P. Arnold, K. Enqvist,
M. Laine, L. Yaffe, and especially G. Moore for discussions
on the electroweak phase diagram; D. Son, for
comments on how particle and anti-particle condensates mix;
D. Rischke, for numerous discussions, especially for 
pointing out the importance of the color adjoint chiral field,
and the effects of instantons with topological charge $Q>1$;
lastly, G. G. De Roux Taylor, for the figures.
This work was supported in part by DOE grant DE-AC02-98CH10886.


\end{document}